\documentstyle[11pt,newpasp,twoside,epsfig]{article}
\begin{document}
\title{The dark matter halo of the gravitational lens galaxy 0047-2808}
\author{Randall B. Wayth}
\affil{School of Physics. University of Melbourne, 3010. Australia}
\author{Rachel L. Webster}
\affil{School of Physics. University of Melbourne, 3010. Australia}
\begin{abstract}
The location of the images in a multiple-image gravitational lens system are strongly dependent on the orientation angle of the mass distribution. As such, we can use the location of the images and the photometric properties of the visible matter to constrain the properties of the dark halo. We apply this to the optical Einstein Ring system 0047-2808 and find that the dark halo is almost spherical and is aligned in the same direction as the stars to within a few degrees.
\end{abstract}
\section{Introduction}
Numerical simulations of Cold Dark Matter (CDM) have been very successful in reproducing the observed large scale structure of the universe. The CDM model predicts that the dark matter (DM) haloes of today's galaxies are assembled through successive mergers of smaller haloes. Simulations using only dark matter predict that the haloes should be quite prolate, however it is not clear how gas and/or stars interacting with the dark matter will change the shape of the halo. Studies have suggested that the DM halo can become more or less cuspy ({El-Zant}, {Shlosman}, \&  {Hoffman}, 2001; {Tissera} \& {Dominguez-Tenreiro}, 1998) and rounder ({Evrard}, {Summers}, \& {Davis}, 1994; {Dubinski}, 1994) after the interaction with stars and gas. An important test of galaxy formation and evolution models will be to compare the shape and profile of galaxy haloes with observed haloes. Thus, simple questions such as: ``Do we expect the visible and dark matter to be aligned in elliptical galaxies?'' and ``Is the dark matter density in the central regions changed by the gravitational dominance of the stars?'' must be answered with observations. For instance: the Milky Way, despite being a spiral galaxy, appears to have an almost spherical halo ({Ibata} {et~al.}, 2001).

Gravitational lensing offers a method to tightly constrain the shape of DM haloes in the population of medium redshift ($0.1 < z < 1.0$) lens galaxies. The image positions in a lens system are highly sensitive to the orientation of the overall mass profile. {Keeton}, {Kochanek}, \&  {Falco} (1998) showed that the \emph{overall} mass distribution is typically aligned with the visible matter using a sample of lens galaxies and a simple SIE mass model. However, depending on the lens galaxy, the stellar mass can contribute a substantial fraction of the total mass inside the image. The extreme case is the lensed QSO 2237+0305 where the dark matter constitutes only 4\% of the projected mass inside the images ({Trott} \& {Webster}, 2002). In this case we expect the visible matter orientation and the total matter orientation derived from a lensing analysis to be very similar. The logical next step is to use a more complicated (stars + halo) model for the lens galaxy to determine the properties of the DM halo alone.

In this paper we use an implementation of the LensMEM algorithm ({Wallington}, {Kochanek}, \&  {Narayan}, 1996) and a stars+halo lens model to study the optical Einstein Ring 0047-2808 ({Warren} {et~al.}, 1999, 1996) using data from the HST. This system is well suited for the study because it is an isolated lens galaxy so we expect any external shear contributions to be small. The system is a $z=0.485$ elliptical which is lensing a background starbursting galaxy at $z=3.6$.

The algorithm we employ performs a non-parametric source reconstruction to match the observed data for a given lens model. The goodness-of-fit of the model is calculated using a $\chi^2$ taking into account the degrees of freedom used in the source. In this paper we assume $H_0=70$ kms$^{-1}$Mpc$^{-1}$ and $(\Omega_m,\Omega_{\lambda})=(0.3,0.7)$.
\section{Method}
\begin{figure}
\plotone{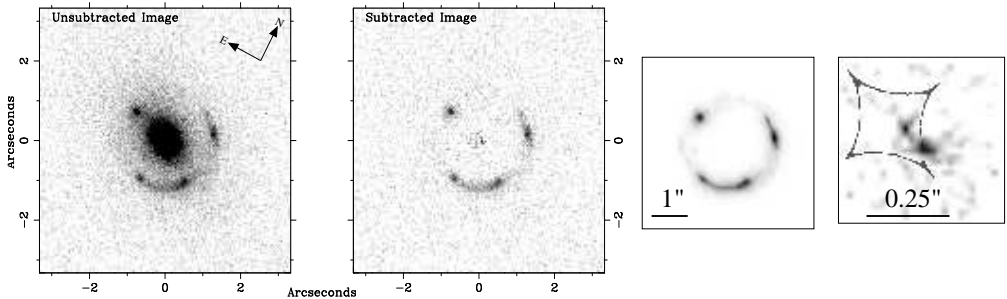}
\caption{Data and model for the 0047-2808 lens. Left: the HST image and image with the galaxy subtracted. Right: Model image and reconstructed source for the PIEP model.}
\label{fig:img_and_model}
\end{figure}

The data was reduced as described in {Wayth} {et~al.} (2002). The final image of the ``ring'' is 133x133 with $0.05\arcsec$ pixels as shown in Figure \ref{fig:img_and_model}.
The lens galaxy was best fit with a Sersic profile, where the surface brightness as a function
of radius $r$ is $\Sigma=\Sigma_{1/2} \exp\{-B(n)\lbrack(r/r_{1/2})^{1/n}-1\rbrack\}$.
The parameter $n$ quantifies the shape of the profile: the values $n=0.5$, $n=1$, and $n=4$ correspond to the Gaussian, exponential, and de Vaucouleurs profiles.
Profiles with larger $n$ are more cuspy. $B(n)$ is a constant for a particular $n$ and we used the series asymptotic
solution for $B(n)$ provided by {Ciotti} \& {Bertin} (1999). Additional parameters used for the light profile are the axis ratio ($q$) and orientation angle ($\theta_s$).
The fitted parameters are shown in Table \ref{tab:phot_fits}.
\begin{table}
\begin{tabular}{lr}
$R_{1/2}$ (pixels) & $21.69$ \\
$\Sigma_{1/2}$ (counts) & $0.7$ \\
$q$ & $0.693$ \\
$\theta_s$ ($\deg$) & $125$ \\
n & $3.115$
\end{tabular}
\caption{Photometric parameters for the lens galaxy.}
\label{tab:phot_fits}
\end{table}

We model the galaxy stellar component with fixed parameters from the photometry and allow only the M/L to vary. The halo is modelled as a Pseudo-Isothermal Elliptic Potential (PIEP) with a finite core. The PIEP model is defined by the lensing potential
$\psi = b[r_c^2 + (1-\epsilon)x^2 + (1+\epsilon)y^2]^{1/2}$ where $r_c$ is the core radius, $b$ is the mass scale (Einstein radius) and $\epsilon$ is the ellipticity.
An additional parameter is used for the orientation angle ($\theta_h$, measured anti-clockwise from horizontal).
It is worth noting that the lens can be fit with the PIEP model alone (without a core) with the parameters $b=1.165, \epsilon=0.08$ and $\theta_h=129$. We use this mass scale for the halo model. The source in this system actually has two distinct components. The two-component model explains the location and brightness of all features in the image with a standard lens model. Figure \ref{fig:img_and_model} shows the model source and corresponding image for the plain PIEP model. 

The mass enclosed inside the image is tightly constrained by the Einstein radius. We use this constraint to normalise the stellar M/L for a halo of a given core radius. A large core is equivalent to a constant M/L mass model, whereas a small core will generate an unrealistically low M/L for the observed stellar component of the lens.

In preliminary tests, we found that we cannot fit the data for $r_c \ga 7\arcsec$ (42kpc physical scale length) i.e. constant M/L models cannot fit the data. Therefore we have restricted our analysis to halo core radii $< 7\arcsec$. For the range of allowed core radius values, we have calculated the range of halo ellipticity and orientation angle which produce acceptable fits to the data.
\section{Results}
\begin{figure}
\plottwo{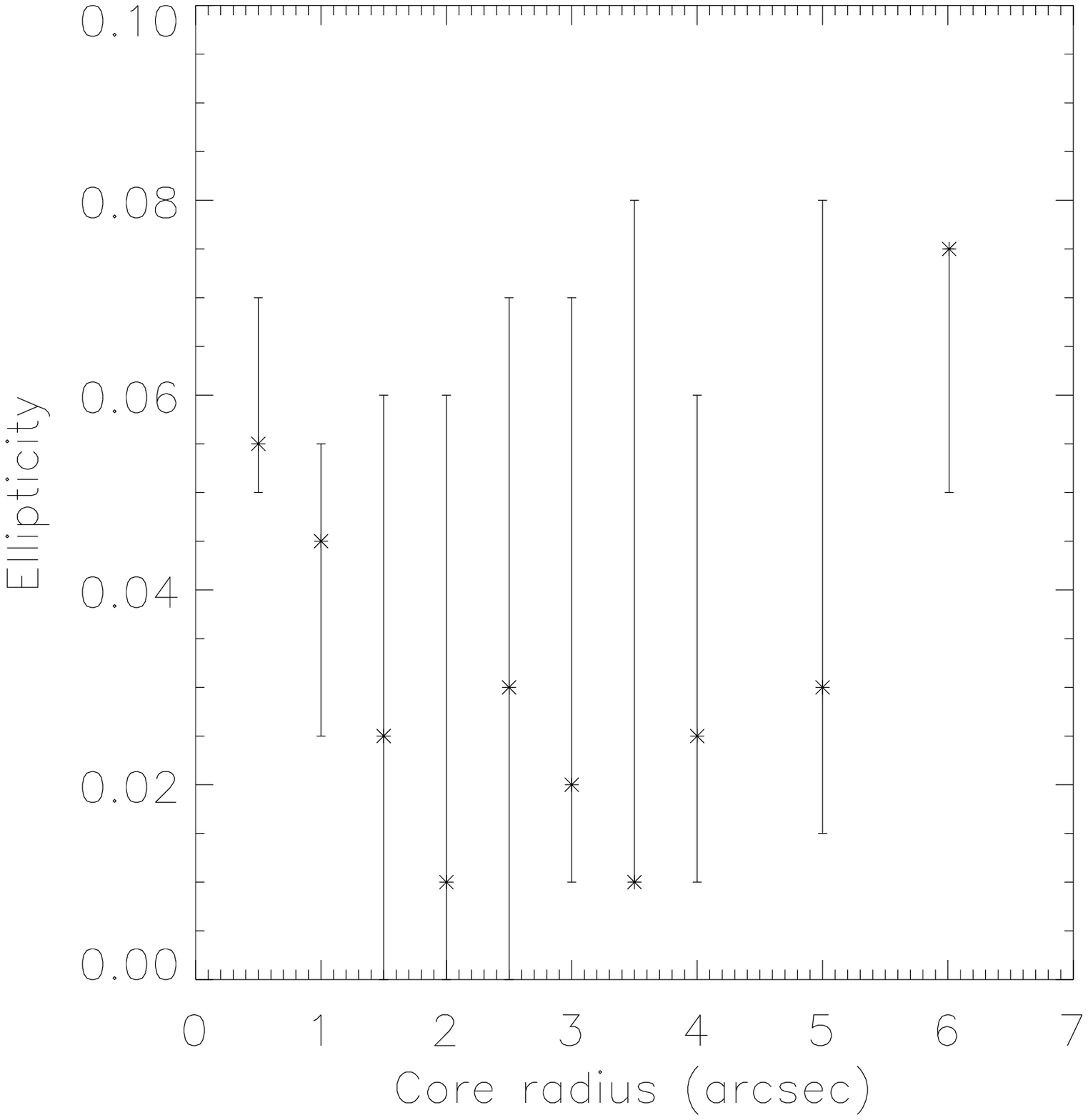}{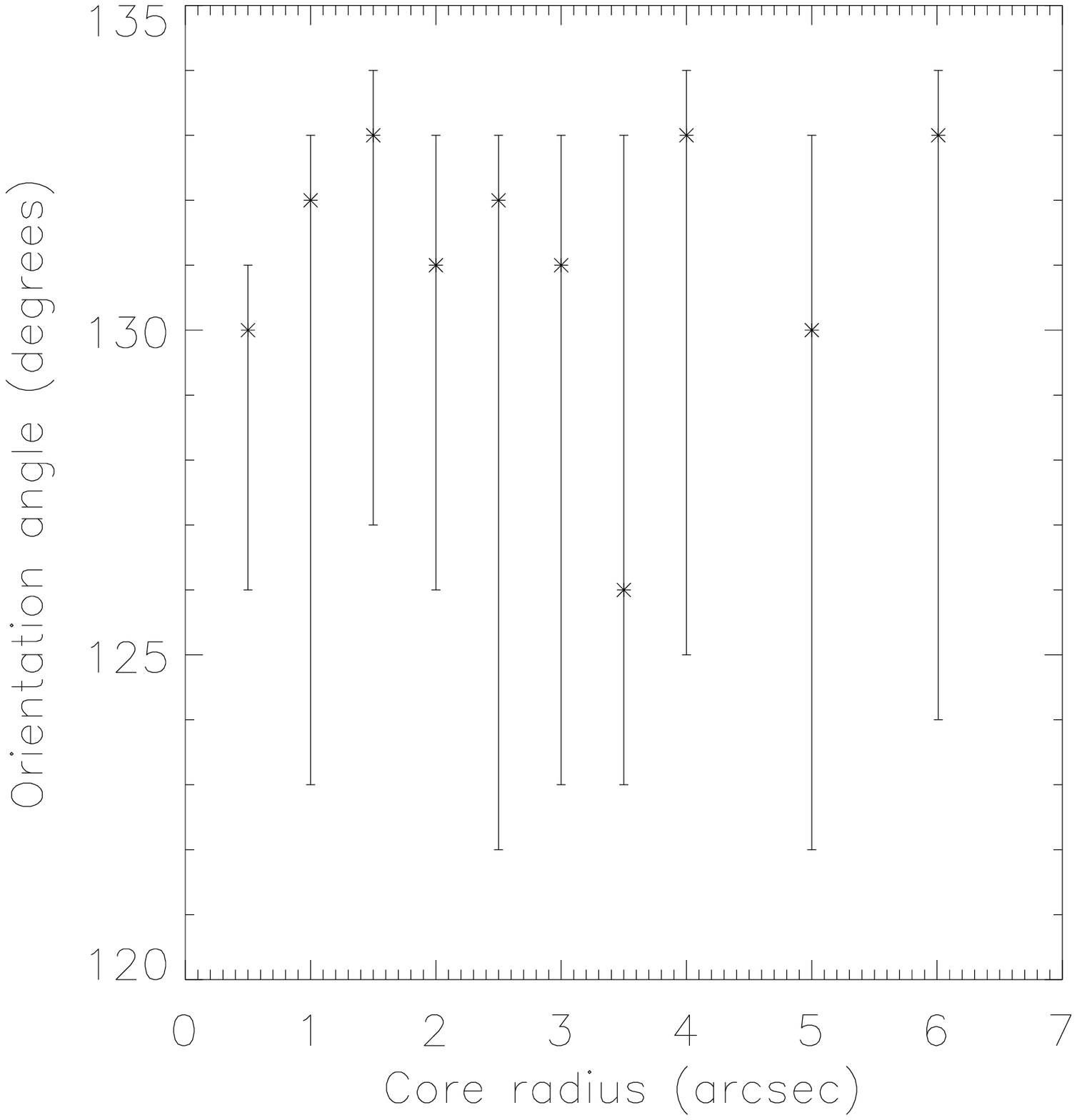}
\caption{Constraints on the dark matter halo. Asterisks indicate the best fitting parameter values. The error bars indicate the range of the parameter which will produce acceptable ($1\sigma)$ $\chi^2$ values. Left: The acceptable range of ellipticity ($\epsilon_h$) for various core radii. Right: the acceptable range of orientation angle ($\theta_h$)}
\label{fig:halores}
\end{figure}

Figure \ref{fig:halores} plots the acceptable range of halo ellipticity and orientation angle as a function of core radius. On the left, we see that the halo ellipticity is consistently less than the stellar ellipticity. For $1.5\arcsec < r_c < 2.5\arcsec$, the data permit a halo with projected mass density which is circular, although in all cases the best solution has a halo with non-zero ellipticity.

On the right of Figure \ref{fig:halores}, the plot shows that the halo orientation angle is independent of the core radius and is in the same direction as the projected stellar major axis (within errors). The acceptable range of orientation angles for $1.5\arcsec < r_c < 2.5\arcsec$ are for non-zero ellipticity.

\section{Conclusion}
By using a lens model which separates the stars from the halo, we have been able to determine some of the basic properties of the dark matter halo in the lens system 0047-2808. We find that the projected ellipticity of the halo is not circular, but is substantially rounder than the observed stellar ellipticity. A small range of halo core radii values ($1.5\arcsec < r_c < 2.5\arcsec$) allow the projected halo mass to be circular.

The halo's core, modelled as a constant density region, must be $< 7\arcsec$ to fit the observation. The core size could be further constrained by applying realistic limits to the stellar M/L which we intend to do in further work.

Finally, we find that although the halo is less elliptical than the stars, the orientation angle of the star's and halo's major axis are the same within errors.

\end{document}